\documentclass[a4paper, 11pt, fleqn]{article}
\usepackage{longtable}
\usepackage{graphicx}
\usepackage{amssymb,amsmath,epsfig,array}

\begin{document}

\title{\bf{Spectrum of the Yellow Symbiotic Star LT Delphini
before, during, and after the 2017 Outburst}}

\author{N.P. Ikonnikova\footnote{E-mail:
ikonnikova@gmail.com}, M.A. Burlak, V.P. Arkhipova, V.F. Esipov}

\date{Sternberg Astronomical Institute,
Moscow State University, Universitetskii pr. 13, Moscow, 119992
Russia}

\renewcommand{\abstractname}{ }

\maketitle

\begin{abstract}

LT Del is a yellow symbiotic system that consists of a bright
K3-type giant and a hot subdwarf with a temperature $\sim 10^5$ K.
We present the results of our spectroscopic observations of LT Del
over the period 2010-2018. In 2017 the star experienced a second
low-amplitude ($\Delta V\sim0.^{m}7$) outburst in the history of
its studies. The emission spectrum of the star represented in the
optical range by hydrogen, neutral and ionized helium lines
underwent significant changes in the outburst. The fluxes in the
HI and HeI emission lines increased by a factor of 5-6, the HeII
$\lambda$4686 line grew by a factor of 10. According to our
estimates, in the 2017 outburst the temperature of the exciting
star rose to $T_{\text{hot}}\sim$130~000 K, while during the first
1994 outburst the change in temperature was insignificant. This
suggests cool and hot outbursts of LT Del by analogy with similar
events of another yellow symbiotic star, AG Dra.
\end{abstract}


\section*{Introduction}

The variable LT Del (Hen 2-467) belongs to a small group of yellow
symbiotic stars whose prototype is generally believed to be  AG
Dra. The system LT Del consists of a bright late-G or early-K
giant and a compact hot star with a temperature $\sim$ 100~000 K.
Allen (1984) classified LT Del as an S-type symbiotic star, a
binary system without a circumstellar dust envelope. The orbital
period of the system is $P_{\textrm{orb}}=476.^{d}0$ (Arkhipova et
al. 2011). Over the entire history of its observations LT Del has
experienced two low-amplitude outbursts, in 1994 and 2017, the
photometric behavior in which was described by Arkhipova et al.
(1995a) and Ikonnikova et al. (2019), respectively.

The spectroscopic observations of LT Del have been carried out for
more than 40 years, beginning with Lutz (1975). The cool component
of the system LT Del has been classified repeatedly. Initially,
Lutz et al. (1976) assigned the spectral type G to the star based
on the presence of the G band and the CaI $\lambda$4226 absorption
line in the spectrum. Subsequently, Munari and Buson (1992)
obtained an estimate of G6III from the infrared (IR) CaII triplet.
Arkhipova et al. (1995a) took the spectral type of the cool
component as G8 II based on photometric data. Yet another
estimate, K3-4, was obtained by 3500--7000~\AA.

Pereira et al. (1998) analyzed a high-resolution optical spectrum
for the yellow component of LT Del and found it to be a metal-poor
K giant with $T_\text{eff}=4400$~K, $\log g=1.8$ and [Fe/H]=--1.1.
The atmosphere of the cool star is enriched in the s-process
elements formed as a result of the evolution of the system's hot
component on the asymptotic giant branch (AGB). The high radial
velocity ($V_r=-106.9$ km s$^{-1}$) together with the low
metallicity suggest that LT Del belongs to the Galactic halo
population.

A study of the ultraviolet (UV) spectrum for LT Del based on IUE
data allowed Munari and Buson (1992) to estimate the temperature
of the hot star, $T_{\text{hot}}=200~000$ K, from the energy
distribution in the range 1300--1900 \AA\ and to obtain a
considerably lower value, $T_{\text{hot}}=70~000$ K, by the
Zanstra method from the HeII $\lambda$1640.

The emission spectrum of the gas component represented mostly by
hydrogen, neutral and ionized helium lines is superimposed on the
spectrum of the cool component in the optical range. Lutz et al.
(1976) were the first to measure the relative intensities of the
HI, HeI, HeII $\lambda$4686 and CII $\lambda$4267. Subsequently,
Lutz (1977) detected a change in the intensity ratio of the HeII
$\lambda$4686 and H$\beta$, lines, while Kaler and Lutz (1980)
found variations of the fluxes in the continuum and H$\beta$,
H$\alpha$ and HeII $\lambda$4686 emission lines. Navarro et al.
(1987) derived the relative emission line intensities in the
optical spectral range and compared their results with the data of
Lutz et al. (1976) and Lutz (1977). Munari and Buson (1992)
measured the absolute emission line fluxes in the range from 3835
to 7065 \AA, studied the emission line variability based on
previously published and their own data, and concluded that the
emission sources in the HI and HeII lines are localized in
different regions. Arkhipova et al. (1995b) analyzed the emission
spectrum of the star in the 1994 outburst. Subsequently, based on
their 1995--2009 observations, Arkhipova et al. (2011) confirmed
that the intensities of the Balmer hydrogen and neutral helium
lines trace the pattern of the system's brightness variations
related to the orbital motion, while the flux in the HeII
$\lambda$4686 line does not change within the measurement error
limits.

In this paper we report the results of our spectroscopic
observations over the period from 2010 to 2018. We were able to
trace the development of the emission spectrum for LT Del before,
during, and after the 2017 outburst and to reveal differences in
the spectrum in the 1994 and 2017 outbursts.

\section*{Spectroscopic Observations}

Systematic spectroscopic observations of LT Del have been carried
out with a 1.25-m telescope at the Crimean Astronomical Station of
the Sternberg Astronomical Institute of the Moscow State
University since 1984. The results of the observations in
1984--2009 were reported previously in Arkhipova et al. (1995b,
2011).

\begin{figure*}
 \includegraphics[scale=2.0]{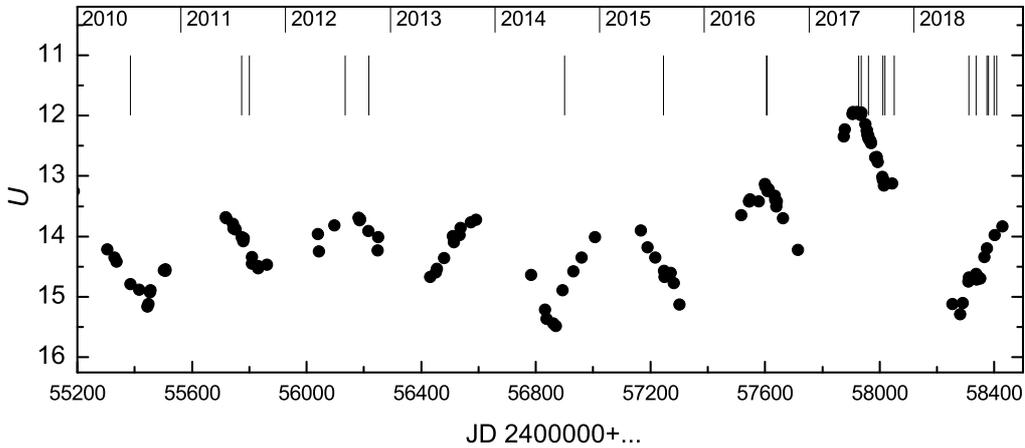}
 \caption{The $U$-band light curve for LT Del over 2010--2018 from Ikonnikova et al. (2019).
 The vertical bars mark the times of spectroscopic observations.}
 \label{Usp}
\end{figure*}

The spectra from 2010 to 2018 were taken at the same telescope
with an ST-402 CCD array and a 600 lines mm$^{-1}$ diffraction
grating, which gave a resolution (FWHM) of about 7.5 \AA. The
investigated spectral range was from 4200 to 7200 \AA\ and, at
some dates, to 9000 \AA. Data on the observational material are
contained in Table~\ref{tab:sp}, while Fig.~\ref{Usp} shows the
times of spectroscopic observations together with the $U$ light
curve from Ikonnikova et al. (2019). Table~\ref{tab:sp} lists the
orbital phases calculated  with the following linear elements:
JD(Min) = 2445930 + 476.$^{d}$0$E$ (Arkhipova et al. 2011). The
stars 50~Boo, 18~Vul, 29~Vul, 40~Cyg, 57~Cyg were used as
standards. The absolute spectral energy distributions for the
standards in the range 4000--7650 \AA\ were taken from the
spectrophotometric catalogue by Glushneva et al. (1998) and were
extended to 9000 \AA\ using data from the atlas of standard
stellar spectra by Pickles (1998). The spectra were reduced using
the standard CCDOPS and MAXIM packages and the SPE code developed
by Sergeev and Heisberger (1993).


\begin{table}
 \caption{Log of spectroscopic observations}
\begin{center}
 \label{tab:sp}
 \begin{tabular}{lcccc}
  \hline
  Date&JD&Phase&Range, \AA \\
  \hline

July 7,  2010&2455385&0.86&4000-7200\\
July 31, 2011&2455774&0.68&4000-7200\\
Aug. 27,  2011&2455801&0.74&4000-7200\\
July 26, 2012&2456135&0.44&4000-7200\\
Oct. 16, 2012&2456217&0.61&4000-7200\\
Aug. 30, 2014&2456900&0.05&4000-7200\\
Aug. 10, 2015&2457245&0.77&4000-7200\\
Aug.  5, 2016&2457606&0.53&4000-9000\\
June 21, 2017&2457926&0.20&4000-9000\\
June 30, 2017&2457935&0.22&4000-9000\\
July 25, 2017&2457960&0.27&4000-9000\\
Sep. 15, 2017&2458012&0.38&4000-9000\\
Sep. 22, 2017&2458019&0.40&4000-9000\\
Oct. 23, 2017&2458050&0.46&4000-7200\\
July 11, 2018&2458311&0.01&4000-9000\\
Aug. 6,  2018&2458337&0.06&4000-9000\\
Sep. 12, 2018&2458374&0.14&4000-9000\\
Sep. 17, 2018&2458379&0.15&4000-9000\\
Oct.  8, 2018&2458400&0.20&4000-9000\\
Oct. 16, 2018&2458408&0.21&4000-9000\\

\hline
 \end{tabular}
\end{center}
\end{table}



\section*{Analysis of the Emission Spectrum for LT Del}

In comparison with the spectra of classical symbiotic stars with
red cool components, the optical emission spectrum of LT Del is
quite poor and is represented only by hydrogen, neutral and
ionized helium lines. There are no forbidden lines, in particular,
[OIII] and [NII] emissions, and no feature characteristic for the
spectra of symbiotic stars, the Raman OVI $\lambda$6825.
Absorption features are clearly seen in the spectrum of LT Del in
quiescence: the MgI $\lambda$5167, DNaI, BaII $\lambda$6496 and
the IR CaII triplet. However, the low spectral resolution does not
allow us to quantitatively analyze the absorption spectrum of the
system's cool component.

We measured the absolute fluxes and equivalent widths of the
H$\gamma$, H$\beta$, H$\alpha$, HeI $\lambda$4388, $\lambda$4713,
$\lambda$4921, $\lambda$5016, $\lambda$5048, $\lambda$5876,
$\lambda$6678, $\lambda$7065, $\lambda$7281, HeII $\lambda$4686,
5412 emission lines in the 2010--2018 spectra and give them in
Tables ~\ref{tab:int} and ~\ref{tab:EW}, respectively. We estimate
the accuracy of measuring the fluxes and equivalent widths to be
within the range from 5\% strong lines to 10 \% for weaker ones.

\begin{figure*}
 \includegraphics[scale=1.7]{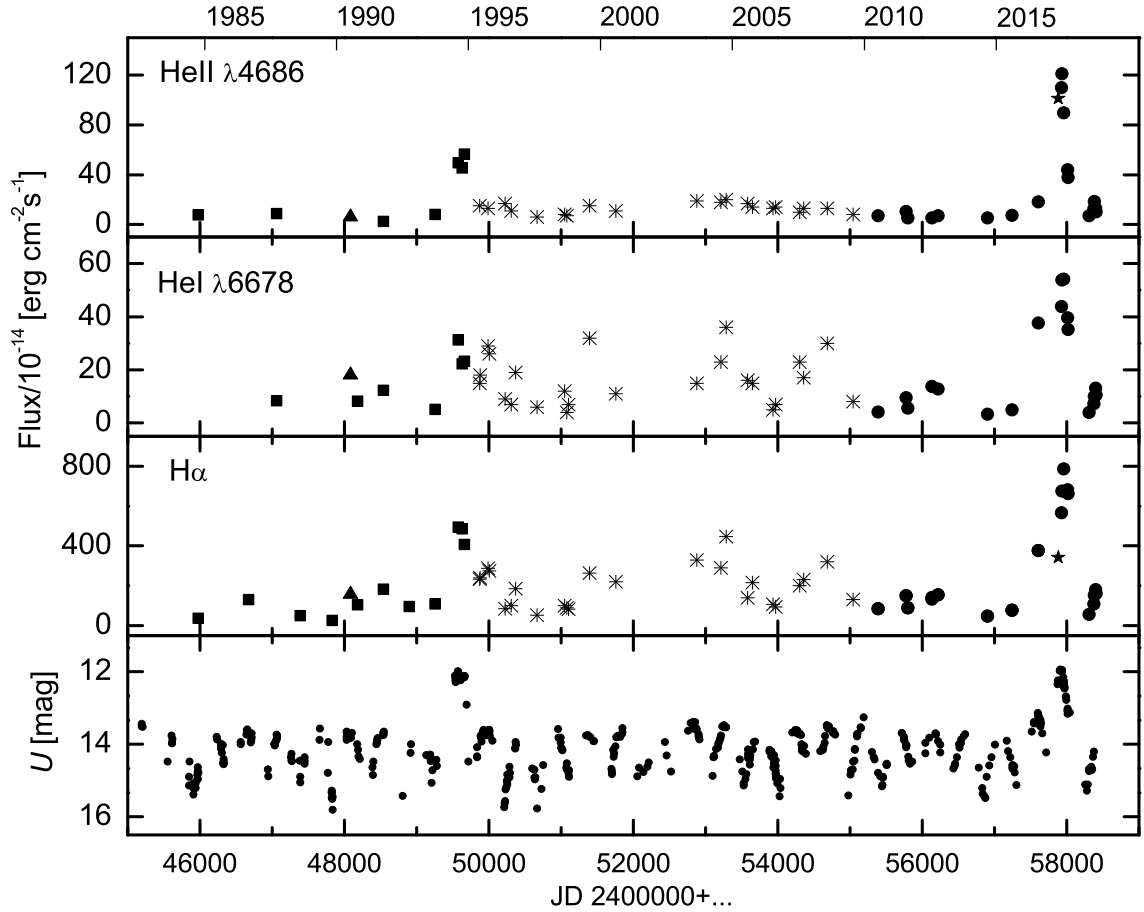}
 \caption{Variations of the fluxes in the H$\alpha$, HeI $\lambda$6678, and HeII $\lambda$4686
 lines with time over the period from 1984 to 2018 according to the data from Arkhipova et al. (1995b)
 (squares), Munari and Buson (1992) (triangles), Arkhipova et al. (2011) (asterisks), Munari et al.
 (2017) (stars), and our new observations (filled circles). The lower panel shows the $U$ band light
 curve for LT Del based on data from Arkhipova et al. (1995a, 2011) and Ikonnikova et al. (2019).}
 \label{flux-jd}
\end{figure*}

The emission line fluxes and equivalent widths change noticeably
with time. The variations in the fluxes of the strongest lines,
H$\alpha$, HeI $\lambda$6678 and HeII $\lambda$4686, over the
period from 1984 to 2018 according to the data from Munari and
Buson (1992), Arkhipova et al. (1995b, 2011), Munari et al.
(2017), and our new observations, along with the $U$ band light
curve from Arkhipova et al. (1995a, 2011) and Ikonnikova et al.
(2019), are shown in Fig.~\ref{flux-jd}.


\begin{figure*}
 \includegraphics[scale=1.5]{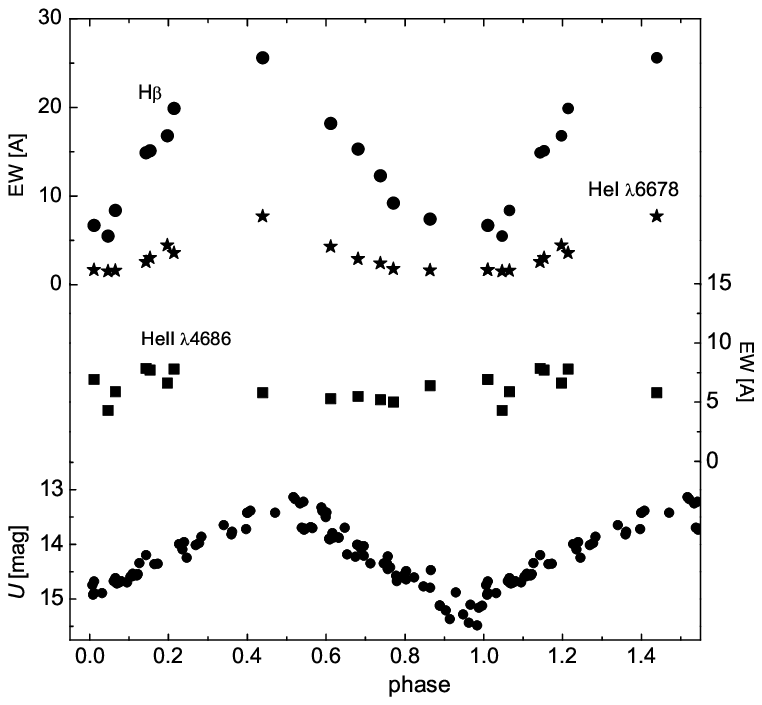}
 \caption{Emission line equivalent width (EW) and $U$ magnitude versus orbital phase.
 The data refer to the quiescent state of LT Del: 2010--2015 and 2018.}
 \label{EW}
\end{figure*}

Outside the outbursts the variability of the HI and HeI emission
lines is synchronized with the $U$ magnitude variations and
reflects a change in the visibility conditions for the line
formation region during the orbital motion. In contrast to
low-excitation lines, the flux in the HeII $\lambda$4686 line does
not depend on the orbital phase, as was shown previously (Munari
and Buson 1992; Arkhipova et al. 2011) and is confirmed by our new
observations. This is direct evidence that no eclipse of the
compact HeII emission line formation region and the hot star
occurs in the system. Figure~\ref{EW} shows the variations in the
equivalent widths of the H$\beta$, HeI $\lambda$6678 and HeII
$\lambda$4686 lines with orbital phase, along with the $U$-band
phase light curve, from the observations in quiescence from 2010
to 2015 and in 2018.


\begin{figure*}
 \includegraphics[scale=1.1]{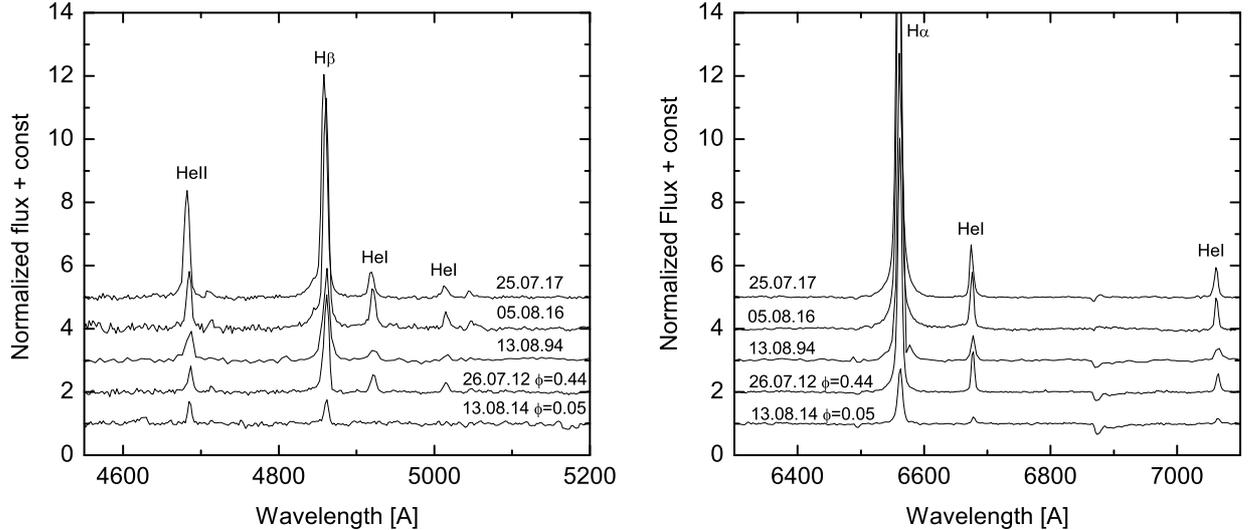}
 \caption{Fragments of the normalized spectrum for LT Del at various orbital phases in quiescence, the 1994 and 2017 outbursts, and the pre-outburst state in 2016.}
 \label{br}
\end{figure*}

In 2016, when the star was in its pre-outburst state, we took a
spectrum near the orbital maximum ($\phi = 0.53$). At this time
the fluxes in the lines, including HeII $\lambda$4686, were
appreciably higher than they were at close phases in preceding
years, as shown in Fig.~\ref{br}, where, in particular, fragments
of the 2016 ($\phi = 0.53$) and 2012 ($\phi = 0.44$) spectra are
presented.

In 1994 the first low-amplitude ($\Delta V \sim 0.^{m}8$) outburst
of LT Del in the history of its studies occurred at orbital phases
0.57--0.90 and lasted $\sim$ 160 days (Arkhipova et al. 1995a).
The absorption features belonging to the cool component remained
visible in the outburst. The fluxes in the HI, HeI, and HeII
emission lines were enhanced by a factor of 2--3 compared to their
mean values for these orbital phases. The temperature of the hot
star in the outburst was estimated by the Zanstra method from the
HeII $\lambda$4686, line to be
$T_\text{hot}(\text{flash})=65~000$~K. This value is lower than
that obtained for the quiescent state of the star,
$T_\text{hot}(\text{quiet})=74~000$~K (Arkhipova et al. 1995b).

Munari et al. (2017) reported a new outburst of LT Del in 2017.
According to the spectroscopic data obtained by them on May 8,
2017, with the Asiago 1.22-m telescope (Italy) (the range
3200--7900 \AA, a dispersion of 2.31 \AA/pixel), the nebular
continuum became stronger and veiled the continuum features of the
cool star. The HeII $\lambda$4686 emission line in this spectrum
was slightly stronger than H$\beta$ ($10.11\times 10^{-13}$ vs
$9.71\times 10^{-13}$ erg cm$^{-2}$ s$^{-1}$), the high-excitation
OIII $\lambda$3444 and $\lambda$3429, OIV $\lambda$3411, [Ne V]
$\lambda$3345 and $\lambda$3427 lines were visible. The flux ratio
of Balmer lines H$\alpha$:H$\beta$:H$\gamma$:H$\delta$ was
3.53:1.00:0.46:0.36.

\begin{figure*}
 \includegraphics[scale=1.5]{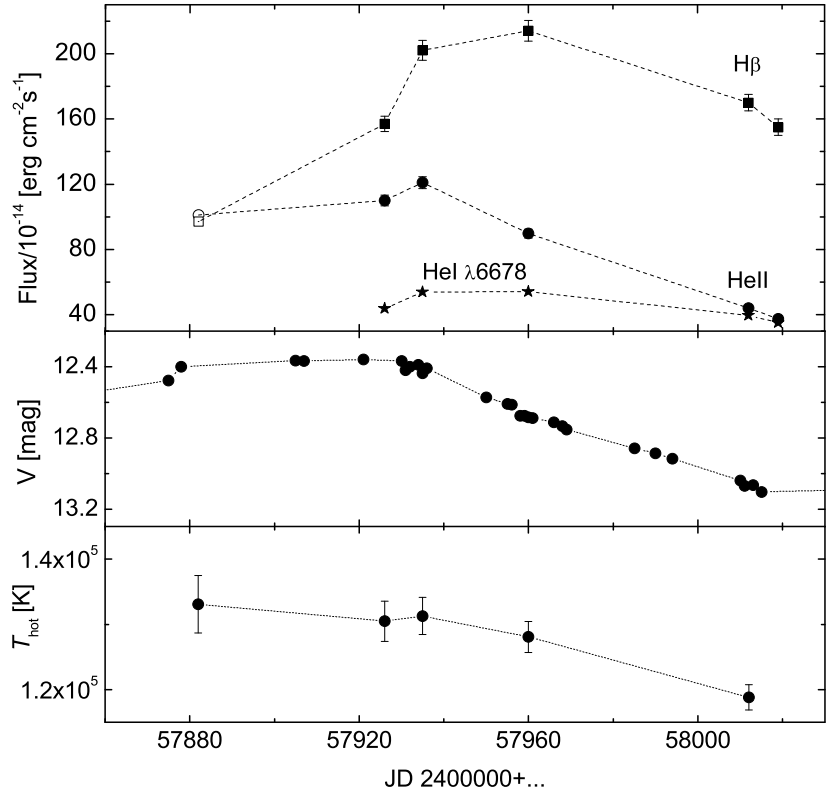}
 \caption{Variations of the absolute fluxes in the H$\beta$, HeI $\lambda$6678, HeII $\lambda$4686 emission lines,
 the $V$ magnitude, and the temperature of the hot star ($T_\text{hot}$) in 2017. The open symbols indicate the observations by Munari et al. (2017).}
 \label{hbheiiV}
\end{figure*}

In 2017, from June 21 to October 23, we took six spectrograms for
LT Del. It can be seen from Table~\ref{tab:int} that all the
emission lines in the outburst strengthened considerably. The
fluxes increased by a factor of 5--6 in the HI and HeI lines and
by a factor of 10 in HeII $\lambda$4686 compared to the quiescent
state of the system at the same orbital phases. The HeI
$\lambda$4388, $\lambda$4713, $\lambda$5048, $\lambda$7281
emission lines, which were not detected previously in the spectrum
of LT Del, appeared. Figure~\ref{hbheiiV} shows the variations of
the fluxes in the H$\beta$, HeI $\lambda$6678 and HeII
$\lambda$4686 lines in 2017 and the $V$ band light curve. In the
spectrum taken on May 8, 2017, (Munari et al. 2017) HeII
$\lambda$4686 was slightly stronger than H$\beta$, while in our
later spectra (on June 21, 2017, and later) H$\beta$ is much more
intense than HeII $\lambda$4686. According to our data, the peak
flux in the HeII $\lambda$4686 line was observed near maximum
light (JD=2457935), while the fluxes in the hydrogen and neutral
helium lines reached their peak values $\sim$30 days later, when
the $V$ brightness of the star decreased by $0.^{m}3$, which may
be related to different localizations of the formation regions of
the ionized helium lines (close to the hot subdwarf) and the
hydrogen and neutral helium emissions (a region in the wind and
near the surface of the cool component). In the outburst the
absorption features belonging to the cool star became
undetectable, because the contribution of the gas continuum and
the hot star increased.

In 2018 the star returned to quiescence. From July 11 to October
16 we took six spectrograms while the system was at orbital phases
from 0.01 to 0.21. As expected, during the egress from the orbital
minimum the fluxes in the HI and HeI lines rose. The HeII
$\lambda$4686 line in 2018 was stronger than before the 2017
outburst, except for the date JD=2458311 referring to orbital
phase 0.01. In 2018 absorptions of the cool star clearly
manifested themselves in the spectrum of LT Del.


\subsection*{The Ratio of the Fluxes in the HeII ?4686 and H$\beta$ Lines and the Temperature of the Hot Star}

\begin{figure*}
 \includegraphics[scale=0.6]{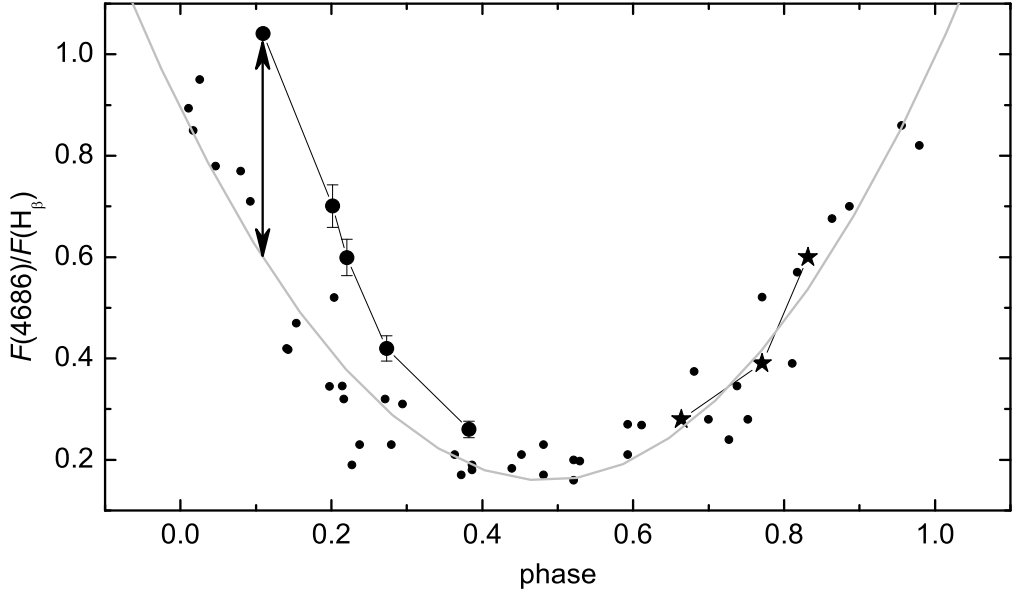}
 \caption{Ratio  $F(4686)/F(\text{H}\beta)$  versus orbital phase. The small circles refer to the quiescent state of LT Del, the gray solid line indicates the fit to these data by a parabola. The stars and big circles describe the 1994 and 2017 outbursts, respectively. The double arrow indicates the difference in $F(4686)/F(\text{H}\beta)$ at one phase in quiescence and in the 2017 outburst.}
 \label{f4686fhb}
\end{figure*}

The ratio of the absolute fluxes in the HeII $\lambda$4686 and
H$\beta$ lines in the spectrum of LT Del changes with orbital
phase; these changes concern H$\beta$, while the intensity of HeII
$\lambda$4686 remains approximately constant.
Figure~\ref{f4686fhb} shows the phase dependence of
$F(4686)/F(\text{H}\beta)$ according to the data from Lutz et al.
(1976), Lutz (1977), Allen (1984), Navarro et al. (1987), Munari
and Buson (1992), Arkhipova et al. (1995b, 2011), Munari et al.
(2017), and our new observations. In quiescence the ratio
$F(4686)/F(\text{H}\beta)$ varies from 0.16 at maximum to 0.95 at
minimum light. At phase 0.5 the HeII and HI emission line
formation regions are completely visible, while at phase 0.0 much
of the hydrogen line emission zone is hidden from the observer by
the cool component.

In the active state the ratio $F(4686)/F(\text{H}\beta)$  also
varies with the phase, as can be seen from Fig.~\ref{f4686fhb}.
However, there is a significant difference between the 1994 and
2017 outbursts: whereas in 1994 the values of
$F(4686)/F(\text{H}\beta)$ were virtually indistinguishable from
those for the quiescent state of the system at the same orbital
phases, in 2017 they were noticeably higher, which may be
indicative of an increase in the temperature of the hot star.


Iijima (1981) proposed a modified method of Ambartsumyan (1932) to
estimate the temperature of the central stars of planetary nebulae
and the hot components of symbiotic binaries in the form of an
equation:


\begin{equation}
\begin{array}{c}

T_\text{hot}=(19.38\sqrt{\frac{2.22F(4686)}{4.16F(\text{H}\beta)+
9.94F(4471)}}+5.13)\times10^{4},

\end{array}
\end{equation}
where $T_\text{hot}$ -- is the temperature of the hot star,
$F(4686)$, $F(\text{H}\beta)$ and $F(4471)$ are the fluxes in the
$\lambda$4686, H$\beta$ and HeI $\lambda$4471 lines corrected for
interstellar extinction.

In the spectrum of LT Del in quiescence HeI $\lambda$4471 HeI is
distinguished with difficulty against the continuum background, in
the outburst it is extremely weak and its measurement is
unreliable. The other lines of the HeI triplet series,
$\lambda$5876 and $\lambda$7065, are fairly strong and, therefore,
we decided to use HeI $\lambda$5876 to take into account the
fraction of photons from the hot star absorbed by helium atoms.
The effective recombination coefficient for HeI $\lambda$5876 was
calculated with the PyNeb package (Luridiana et al. 2015), which
uses the results of Porter et al. (2012, 2013) to compute the HeI
atom. Given what has been said above, we transformed (1) to the
following equation:


\begin{equation}
\begin{array}{c}

T_\text{hot}=(19.38\sqrt{\frac{2.22F(4686)}{4.16F(\text{H}\beta)+
2.88F(5876)}}+5.13)\times10^{4},

\end{array}
\end{equation}
where $F(4686)$, $F(\text{H}\beta)$ and $F(5876)$ are the fluxes
in HeII $\lambda$4686, H$\beta$ and HeI $\lambda$5876, and used it
to estimate the temperature of the exciting star in LT Del. The
line fluxes were corrected for interstellar reddening with
$E(B-V)=0.^{m}2$ (Skopal 1995).

The most reliable value of $T_\text{hot}$ for LT~Del can be
obtained for orbital phase 0.5, when the region of ionized
hydrogen is most open to the observer. For other configurations of
the system part of the HII region is covered by the cool
component, which increases the ratio $T_\text{hot}$. Using Eq.~2
we found that in quiescence at maximum light
$T_\text{hot}(\text{quiet})\sim 105~000$ K, which is in agreement
with the estimate of $T_\text{hot}$ = 100~000 K from Skopal (2005)
and considerably higher than the temperature of the hot star
$T_\text{hot}=70~000$ K inferred by the Zanstra method from HeII
$\lambda$1640 (Munari and Buson 1992).

We also estimated the temperature of the exciting star in the 2017
outburst. Our spectroscopic observations in 2017 refer to the
times when part of the HI and HeI emission region was covered by
the disk of the cool star. We assumed that in the outburst the
sizes of the HII and HeII regions did not change and they were
obscured by the disk of the cool star during the orbital motion in
the same way as in quiescence, while the HeIII region was always
visible.

According to our estimates, the temperature of the hot star in the
2017 outburst increased to$\sim$130~000 K, remaining at this level
until JD$\sim$2457935, after which it began to decrease together
with the fading of the system. Figure~\ref{hbheiiV} shows a plot
of the variations $T_\text{hot}(\text{flash})$ from the 2017
observations.

\subsection*{The Optical Spectral Energy Distribution for LT Del}

\begin{figure*}
 \includegraphics[scale=1.5]{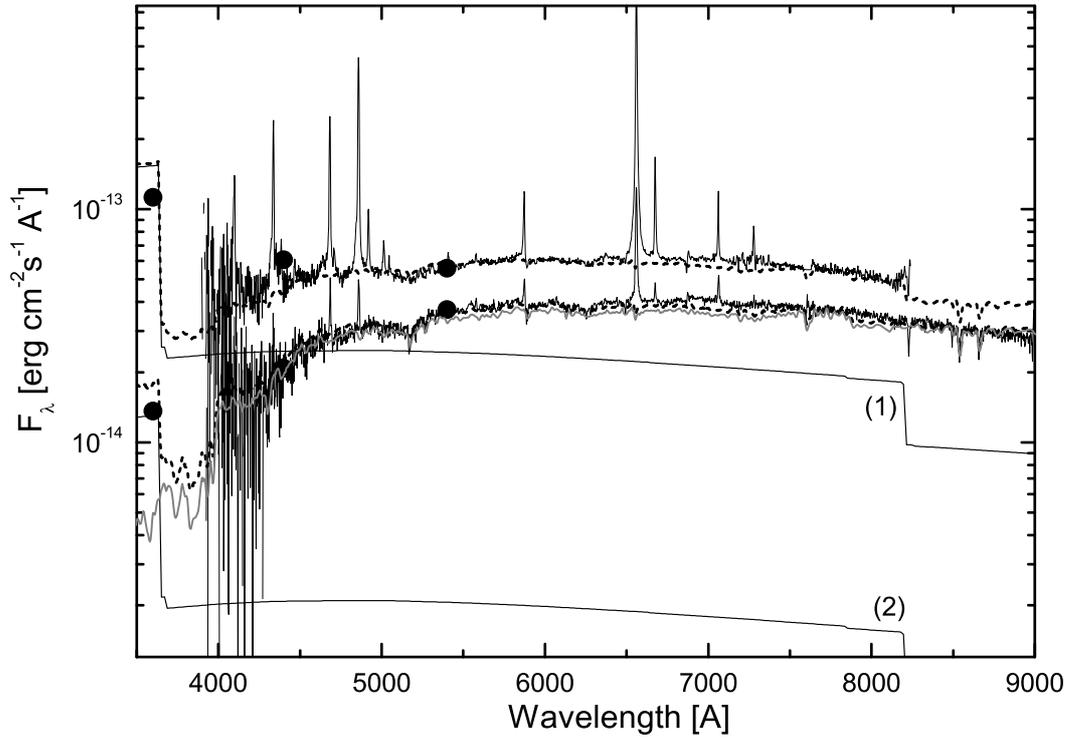}
 \caption{Optical spectra of LT Del corrected for interstellar reddening in the 2017 outburst (July 27, 2017, $\phi=0.27$; the upper spectrum) and in quiescence (July 11, 2018, $\phi=0.01$; the lower spectrum). The dashed lines indicate the model energy distribution curves constructed by adding a constant contributor, a K3-type giant (gray solid line), and a variable gas continuum with $T_{e}=15~000$ K:(1) for the spectrum in the outburst and (2) for the spectrum at minimum light. The circles indicate the fluxes converted from the $UBV$ magnitudes for the corresponding date.}
 \label{model}
\end{figure*}

Figure~\ref{model} presents the spectra corrected for reddening
with $E(B-V)=0.^{m}2$ (Skopal 2005) in the 2017 outburst and in
quiescence at minimum light, along with the model spectra and
$UBV$ photometry.

The model curves are the sums of the emissions from the cool
component and the photoionized nebula absorbing all Lc photons.
The K3 III spectrum was taken from the library of stellar spectra
by Pickles (1998).

The gas continuum was obtained as a sum of the hydrogen and helium
continua with an electron temperature $T_{e}=15~000$ K. The helium
abundance was assumed to be He/H=0.12.

The spectrum taken on July 11, 2018, at an orbital phase close to
minimum light($\phi=0.01$), in the wavelength range 4000--9000
\AA\ is well represented by the emission from a K3-type giant with
the addition of a gas continuum with $T_e=15~000$ K (curve (2) in
Fig.~\ref{model}). The contribution of the emission from the red
giant at $\lambda$5500 exceeds the fraction of the gas continuum
emission by a factor of 16.

In the 2017 outburst the optical continuum level for LT Del rose
significantly due to an increase in the fraction of the emission
from the gas component. In the wavelength range 4000--9000 \AA\
the spectrum of LT Del taken on July 27, 2017, can be represented
by a sum of the emissions from a K3 giant and a gas continuum
(curve (1) in Fig.~\ref{model}) with
$F_{\lambda}(\text{K3})/F_{\lambda}(\text{gas})=1.4$ at
$\lambda$=5500 \AA.

Thus, the gas continuum level in the outburst rose by a factor of
$\sim$10 compared to the quiescent state at an orbital phase close
to 0.0.

Unfortunately, the spectral range accessible to us for analysis
does not allow information about the emission from the hot star to
be obtained.


\subsection*{LT~Del è AG~Dra}

LT Del belongs to a small group of yellow symbiotic systems whose
cool components have a spectra type no later than K5. AG Dra is
generally believed to be the prototype of this group.

AG~Dra and LT~Del a similar set of optical emission lines (HI,
HeI, HeII). The main difference is the presence of the Raman OVI
$\lambda$6825 line in the spectrum of AG Dra, which is absent in
the spectrum of LT Del. A study of the spectrum for AG Dra showed
that this star, just as LT Del, in quiescence exhibits variations
in the equivalent widths of low excitation (HI and HeI) emission
lines with orbital phase and there is no such dependence for the
high excitation HeII $\lambda$4686 (Leedj\"{a}rv et al., 2016). It
was concluded that the HI and HeI lines originate in an extended
gaseous volume together with the continuum emission in the near-UV
and optical spectral ranges, while the HeII lines arise in a
considerably smaller region near the hot star, which is also true
for LT Del.

In outbursts the emission lines in the spectrum of AG Dra
strengthen, but there is no unambiguous relation between their
equivalent widths and and the brightness of the star. For example,
during the active state of AG Dra in 2007 (event F1), when the
brightness of the star rose by $\sim 3.^{m}7$ in the $U$ band, the
H$\alpha$, H$\beta$, HeI $\lambda$6678 and HeII $\lambda$4686
emission lines did not change, while the Raman OVI $\lambda$6825
line weakened compared to the quiescent state. At the same time,
in the 2005 outburst (event E10) with a lower amplitude all the
emission lines in the spectrum of AG Dra strengthened considerably
(Leedj\"{a}rv et al., 2016). This points to a difference between
the outbursts: the temperature of the exciting star rises in some
outbursts and apparently decreases in other ones. Something of
this kind is also observed in LT Del, in which the emission
spectrum changed differently in its outbursts with an
approximately equal amplitude in the $V$ band. In 1994 the
emission line fluxes rose insignificantly, the ratio of HeII
$\lambda$4686  and H$\beta$ did not change compared to the
quiescent state and, therefore, it can be assumed that the
temperature of the hot star did not increase. In 2017 a more
significant change in the emission spectrum was observed: a
considerable strengthening of the emission lines, an increase in
the intensity ratio of the HeII $\lambda$4686 and H$\beta$
emission lines, and a rise of the nebular continuum level,
suggesting an increase in the temperature of the hot source.

\section*{Conclusions}

We presented the results of our spectroscopic observations of the
yellow symbiotic star LT Del.

(1) An analysis of our new data, along with the previously
published ones, showed that in quiescence, between the 1994 and
2017 outbursts, the change in the HI and HeI emission lines was
related to the visibility conditions for the line formation
regions in the binary system consisting of a hot subdwarf and a
cool bright giant and is synchronized with the photometric
variability of the star in the $U$ band. The intensity of HeII
$\lambda$4686 does not depend on the orbital phase, but it showed
amonotonic change: a slight strengthening from 1995 to 2003 and
then a weakening until 2015.

(2) In the pre-outburst state in 2016, at orbital phase 0.53, we
took one spectrum in which all lines strengthened compared to the
observations at close phases in quiescence.

(3) In 2017, during the second outburst of LT Del in the history
of its studies, the emission line fluxes rose significantly. The
HeI $\lambda$4388, $\lambda$4713, $\lambda$5048 and $\lambda$7281
emission lines, which have not been observed previously, have been
detected for the first time in the spectrum. The intensity of HeII
$\lambda$4686 increased by a factor of 10, while the HI and HeI
emission lines strengthened by a factor of 5--6. In the optical
range the contribution of the gas continuum increased, which led
to a weakening of the absorption features in the spectrum of the
cool component.

(4) We estimated the temperature of the hot subdwarf by an
analytical method using the intensities of HeII $\lambda$4686,
H$\beta$ and HeI $\lambda$5876. We showed that, in comparison with
the quiescent state with $T_\text{hot}(\text{quiet})\sim 105~000$
K the temperature in the 2017 outburst rose to
$T_\text{hot}(\text{flash})\sim 130~000$ K, while in the 1994
outburst the change in temperature was insignificant. Therefore,
it can be concluded that LT Del experiences cool and hot
outbursts, as does the prototype of the subclass of yellow
symbiotic stars AG Dra.

(5) In 2018 the star returned to quiescence, but the emission line
fluxes remained higher than those before the 2017 outburst.

(6) We modeled the optical spectrum of LT Del at minimum light and
in the 2017 outburst. The optical spectral energy distribution for
the system was shown to be satisfactorily fitted by a sum of the
emissions from two components: a cool K3 III giant and a gas
continuum with $T_\text{e}=15~000$ K. The gas continuum level in
the outburst was found to have increased by a factor of $\sim$10
compared to the quiescent state at an orbital phase close to 0.0.

\bigskip
REFERENCES
\bigskip

\begin{enumerate}

\item D.A. Allen, Proc. Astron. Soc. Austral. {\bf 5}, 369 (1984).

\item V.A. Ambartsumyan, Circ. Glav. Astron. Obs., {\bf 4}, 8
(1932).

\item V.P. Arkhipova, N.P. Ikonnikova, and R.I. Noskova, Astron.
Lett. $\bf{21}$, 379 (1995a).

\item V.P. Arkhipova, V.F. Esipov, and N.P. Ikonnikova, Astron.
Lett. {\bf 21}, 439 (1995b).

\item V.P. Arkhipova, V.F. Esipov, N.P. Ikonnikova, G.V.
Komissarova, and R.I. Noskova,
      Astron. Lett. {\bf 37}, 377 (2011).

\item I.N. Glushneva, V.T. Doroshenko, T.S. Fetisova, T.S.
Khruzina, E.A. Kolotilov, L.V. Mossakovskaya, S.L. Ovchinnikov,
I.B.Voloshina, VizieR Online Data Catalog III/208 (1998).

\item T. Iijima, 1981, in Carling E. B., Kopal Z., eds,
Photometric and Spectroscopic Binary Systems. Kluwer, Dordrecht,
p. 517

\item N. P. Ikonnikova, G. V. Komissarova, and V. P. Arkhipova,
Astron. Lett. (2019, in press).

\item J.B. Kaler, J.H. Lutz, PASP {\bf 92}, 81 (1980).

\item L. Leedj\"{a}rv, R. G\'{a}lis, L. Hric, J. Merc and M.
Burmeister, MNRAS {\bf 456}, 2558 (2016).

\item V. Luridiana, C. Morisset, R.A. Shaw, Astron. Astrophys.
{\bf 573}, 42 (2015).

\item J.H. Lutz, Bulletin of the American Astronomical Society,
{\bf 7}, 243 (1975).

\item J.H. Lutz, T.E. Lutz, J.B. Kaler,  D.E. Osterbrock, S.A.
Gregory,  Astrophys. J. {\bf 203}, 481 (1976).

\item J.H. Lutz, Astron. and Astrophys., {\bf 60}, 93 (1977).

\item U. Munari, L.M. Buson, Astron. Astrophys. {\bf 255}, 158
(1992).

\item U. Munari, P. Ochner, S. Dallaporta and R. Belligoli,
Astronomer's Telegram, 10361 (2017).

\item S. Navarro, R. Ñostero, P.G. Serrano, L. Carrasco, Rev. Mex.
Astrofis. {\bf 14}, 339 (1987).

\item C.B. Pereira, V.V. Smith, K. Cunha, Astron. J. {\bf 116},
1997, (1998).

\item A.J. Pickles, PASP  {\bf 110}, 863 (1998).

\item R.L. Porter, G.J. Ferland, P.J. Storey, M.J. Detisch, Mon.
Not. Royal Astron. Soc. {\bf 425}, L28 (2012).

\item R.L. Porter, G.J. Ferland, P.J. Storey, M.J. Detisch, Mon.
Not. Royal Astron. Soc. {\bf 433}, L89 (2013).

\item S.G. Sergeev and F. Heisberger, A Users Manual for SPE. Wien
(1993).

\item A. Skopal, Astron. and Astrophys. {\bf 440}, 995 (2005).

\end{enumerate}

\begin{table}

\caption{Observed emission line fluxes in units of $10^{-14}$ erg
cm$^{-2}$s$^{-1}$} \label{tab:int}
\begin{tabular}{|c|c|c|c|c|c|c|c|c|}
\hline

JD&ôàçà&H$\gamma$&HeI&HeII&HeI&H$\beta$&HeI&HeI\\
&&4340&4388&4686&4713&4861&4921&5016\\
\hline

2455385&0.86&3.7&--&7.2&--& 10.2&--&--\\
2455774&0.68&16&--&11.3&--&50.5&4.3&--\\
2455801&0.74&7.5&--&6.3&--&16.8&1.3&--\\
2456217&0.61&10.8&--&7.0&--&28.1&4.7&--\\
2456900&0.05&3.8&--&5.2&--&6.4&--&--\\
2457245&0.77&5.4&--&7.3&--&14.6&--&--\\
2457606&0.53&28.1&3.9&18.1&1.9&91.7&14.1&5.2\\
2457926&0.20&60.1&5.8&110.0&4.3&157.0&18.0&10.7\\
2457935&0.22&82.5&8.1&121.0&7.1&202.0&26.1&13.3\\
2457960&0.27&74.1&7.3&89.8&4.6&214.0&20.5&9.6\\
2458012&0.38&65.6&--&44.2&4.1&170.0&12.5&5.2\\
2458019&0.40&51.5&--&37.6&3.1&155.0&9.7&4.7\\
2458311&0.01&--&--&7.2&--&8.1&--&--\\
2458374&0.14&--&--&12.6&--&30.2&4.7&--\\
2458379&0.15&--&--&18.5&--&39.4&5.3&--\\
2458400&0.20&21.0&--&13.0&--&38.0&4.8&2.1\\
2458408&0.21&--&--&10.2&--&29.5&4.2&2.1\\

\hline \hline
JD&ôàçà&HeI&HeII&HeI&H$\alpha$&HeI&HeI&HeI\\
&&5048&5411&5876&6563&6678&7065&7281\\
\hline
2455385&0.86&--&--&4.5&84.3&4.5&5.3&--\\
2455774&0.68&--&--&6.7&150&9.7&5.8&--\\
2455801&0.74&--&--&2.3&89.3&5.5&3.1&--\\
2456217&0.61&--&--&7.2&154&13.8&6.2&--\\
2456900&0.05&--&--&2.6&47.8&4.3&3.3&--\\
2457245&0.77&--&--&3.7&77.0&--&3.4&--\\
2457606&0.53&--&--&18.8&376.0&37.6&20.0&12.1\\
2457926&0.20&--&4.2&27.1&567.0&43.9&23.6&10.0\\
2457935&0.22&6.4&7.7&31.4&676.0&53.9&28.3&12.5\\
2457960&0.27&3.4&4.4&34.8&786.0&54.2&31.4&11.8\\
2458012&0.38&1.8&2.9&31.8&681.0&39.6&32.5&11.6\\
2458019&0.40&--&2.4&--&662.0&35.2&30.1&8.4\\
2458311&0.01&--&--&4.7&56.1&4.0&5.9&--\\
2458374&0.14&--&--&5.1&109.8&7.2&3.3&--\\
2458379&0.15&--&--&7.2&150.0&10.0&4.7&--\\
2458400&0.20&--&--&6.7&181.0&13.2&7.2&3.2\\
2458408&0.21&--&--&7.1&160.7&10.6&6.6&--\\

\hline
\end{tabular}

\end{table}

\begin{table}

\caption{Emission line equivalent widths ($EW$), \AA}
\label{tab:EW}
\begin{tabular}{|c|c|c|c|c|c|c|c|c|}
\hline

JD&ôàçà&H$\gamma$&HeI&HeII&HeI&H$\beta$&HeI&HeI\\
&&4340&4388&4686&4713&4861&4921&5016\\
\hline

2455385&0.86&4.4&--&6.4&--&7.4&--&--\\
2455774&0.68&8.7&--&5.5&--&15.&2.&--\\
2455801&0.74&7.8&--&5.2&--&12.&1.2&--\\
2456135&0.44&15.9&-&5.8&--&25.6&4.8&2.5\\
2456217&0.61&9.9&--&5.3&--&18.2&3.1&1.9\\
2456900&0.05&3.6&--&4.3&--&5.5&--&--\\
2457245&0.77&6.9&--&5.0&--&9.2&0.8&--\\
2457606&0.53&27.7&1.6&13.2&1.2&62.2&9.7&3.4\\
2457926&0.20&19.9&0.8&34.0&1.3&46.1&5.2&3.3\\
2457935&0.22&24.8&0.4&30.0&1.1&52.3&6.7&3.2\\
2457960&0.27&27.5&1.1&30.0&1.8&71.3&7.1&4.1\\
2458012&0.38&47.3&--&25.5&2.6&90.3&7.1&3.1\\
2458019&0.40&32.4&--&20.8&1.6&71.4&5.2&2.3\\
2458050&0.46&28.4&--&13.1&--&62.3&5.7&3.0\\
2458311&0.01&--&--&6.9&--&6.7&--&--\\
2458337&0.07&--&--&5.9&--&8.4&--&--\\
2458374&0.14&--&--&7.8&--&14.9&--&--\\
2458379&0.15&7.7&--&7.7&--&15.1&1.9&--\\
2458400&0.20&11.2&--&6.6&--&16.8&1.8&0.9\\
2458408&0.21&--&--&7.8&--&19.9&2.9&1.1\\

 \hline \hline
JD&ôàçà&HeI&HeII&HeI&H$\alpha$&HeI&HeI&HeI\\
&&5048&5411&5876&6563&6678&7065&7281\\
\hline

2455385&0.86&--&--&2.1&28.5&1.6&1.6&--\\
2455774&0.68&--&--&2.0&49.2&2.9&1.8&--\\
2455801&0.74&--&--&1.8&41.0&2.4&1.2&--\\
2456135&0.44&--&--&5.1&73.4&7.7&3.9&--\\
2456217&0.61&--&--&1.8&47.4&4.3&2.2&--\\
2456900&0.05&--&--&1.1&19.2&1.5&1.2&--\\
2457245&0.77&--&--&1.4&29.0&1.8&1.2&--\\
2457606&0.53&--&--&8.6&164.0&11.7&7.6&3.0\\
2457926&0.20&1.6&1.7&6.1&124.0&10.4&4.4&1.0\\
2457935&0.22&1.5&2.2&6.8&134.0&11.6&5.6&2.6\\
2457960&0.27&1.5&1.3&8.8&176.0&12.3&7.1&2.7\\
2458012&0.38&0.9&1.5&11.4&249.0&14.1&11.2&3.2\\
2458019&0.40&--&1.0&--&226.0&12.7&10.5&2.8\\
2458050&0.46&--&--&6.7&173.0&10.3&7.8&2.2\\
2458311&0.01&--&--&2.3&24.0&1.6&2.2&--\\
2458337&0.07&--&--&0.8&25.6&1.6&1.0&--\\
2458374&0.14&--&--&1.6&40.9&2.6&1.2&--\\
2458379&0.15&--&--&1.6&45.4&3.0&1.5&--\\
2458400&0.20&--&--&2.4&56.4&4.4&2.3&0.7\\
2458408&0.21&--&--&2.7&56.5&3.6&2.92&--\\

\hline
\end{tabular}

\end{table}

\end{document}